\documentclass[aip,
jcp,
amsmath,amssymb,
preprint,
reprint,
]{revtex4-2}

\usepackage{graphicx}

\usepackage[utf8]{inputenc}   
\usepackage[T1]{fontenc}      
\usepackage[english]{babel}
\usepackage{csquotes}
\usepackage{mathtools}
\usepackage{siunitx}
\usepackage{upgreek}
\usepackage{hyperref}

\usepackage{newunicodechar}   

\DeclareUnicodeCharacter{2013}{--}   
\DeclareUnicodeCharacter{2212}{-}    
\DeclareUnicodeCharacter{0394}{$\Delta$}

\newunicodechar{₂}{$_2$}       
\newunicodechar{₃}{$_3$}       
\newunicodechar{₅}{$_5$}       
\newunicodechar{ }{\,}         
\newunicodechar{μ}{$\mu$} 
\newcommand{\upe}{\text{e}}

\begin{document}


\title{Neural Wavefunction Calculations of $\upmu$SR Spectra with Quantum Muons and Protons} 



\author{Jamie Carr}

\author{Mathias Volkai}

\author{W. M. C. Foulkes}

\author{Andres Perez Fadon}


\affiliation{Department of Physics, Imperial College London, London SW7 2AZ, United Kingdom}

\date{\today}

\begin{abstract}

Accurate prediction of muon hyperfine constants is useful for interpreting muon spin spectroscopy data, yet standard methods such as density functional theory (DFT) compute muon-electron pair density functions, and thus hyperfine constants, by treating the muon as a fixed classical particle. This work uses the variational quantum Monte Carlo method with neural-network trial wavefunctions, a highly accurate and flexible approach recently applied to other quantum chemical problems. The muon can be treated classically or included in the many-particle electron-muon wavefunction, in which case the fully quantum mechanical pair density is obtained directly. We calculate muon hyperfine constants in muoniated methyl and ethyl radicals for both quantum mechanical and fixed classical muons.
The hyperfine constants obtained from our fixed-muon calculations in the methyl and ethyl radicals differ from the corresponding DFT results significantly,
highlighting the limitations of DFT even when the muon is treated classically. The results with quantum muons are closer to experiment after accounting for environmental effects. These findings suggest that explicitly calculating the quantum mechanical muon-electron pair density improves the accuracy of muon hyperfine constant predictions.

\end{abstract}

\maketitle 

\section{Introduction}


In a typical $\upmu$SR experiment, low-energy spin-polarized anti-muons (referred to as ``muons'' in most of the literature and throughout the rest of this paper) of charge $+e$ are implanted into solids and molecules, where they act as sensitive probes of the local magnetic environment.\cite{amato_introduction_2024, rhodes_muoniumsecond_2012, blundell_muon_2021} The angular distribution of the positrons produced by the resulting muon decay reveals details about the local spin density of materials.\cite{amato_introduction_2024, blundell_muon_2021} Frequently, the implanted muon interacts with an electron, causing the many-particle wavefunction to develop a muonium-like component: the muon binds to an electron to produce a Muonium (Mu) atom, chemically analogous to a light hydrogen atom.\cite{hillier_muon_2022, amato_introduction_2024} When a muonium atom binds to an unsaturated closed-shell molecule, the result is a muoniated radical with an unpaired electron and non-zero electronic and muonic spin densities.\cite{hillier_muon_2022, mckenzie_positive_2013, rhodes_muoniumsecond_2012} For such radicals, key measurable quantities are the muon hyperfine constants, which represent the strength of the hyperfine interaction between the muon and electron spins.\cite{hillier_muon_2022, amato_introduction_2024, rhodes_muoniumsecond_2012, blundell_muon_2021} Computational predictions of these muon hyperfine constants can then be compared with experimental $\upmu$SR spectra to identify which chemical species are present in the sample.\cite{walsgrove_probing_2024} 

Hyperfine constants can be measured using transverse field $\upmu$SR (TF-$\upmu$SR).\cite{roduner_muonium_1981, roduner_muonium-substituted_1978, rhodes_muoniumsecond_2012, hillier_muon_2022} In an applied transverse magnetic field, the muon and unpaired electron in a muoniated radical form a superposition of four spin eigenstates: $|\uparrow_\upe \uparrow_\upmu \rangle$, $|\uparrow_\upe \downarrow_\upmu \rangle$, $|\downarrow_\upe \uparrow_\upmu \rangle$, and $|\downarrow_\upe \downarrow_\upmu \rangle$.\cite{hillier_muon_2022, amato_introduction_2024} As the magnetic moment of the electron is much larger than that of the muon, the Zeeman interaction separates these into electron-up and electron-down pairs, while the muon Zeeman and hyperfine interactions determine the energy splitting between the muon spin states within each pair.\cite{hillier_muon_2022, amato_introduction_2024} These splittings manifest themselves in the Larmor precession of the muon spin as two distinct frequencies, which can be measured from the angular distribution of the decay positrons, emitted preferentially along the muon spin direction.\cite{hillier_muon_2022, amato_introduction_2024} As the hyperfine energy shift arises from the muon and electron spins being aligned or anti-aligned with the external magnetic field, the difference between the two precession frequencies cancels the muon Zeeman term and yields the isotropic muon hyperfine constant, $A_\upmu$.\cite{amato_introduction_2024, hillier_muon_2022}


Computational quantum chemistry methods are often used to calculate muon hyperfine constants.\cite{walsgrove_probing_2024, thomas_hyperfine_2006, yamada_accurate_2014,amato_introduction_2024, peck_simulating_2015} While high-accuracy methods such as coupled cluster can describe the ground state reliably, their computational cost is often prohibitive for large systems.\cite{pfau_ab_2020} Density functional theory (DFT) provides a much faster alternative and, although generally less accurate, yields surprisingly good results.\cite{yamada_accurate_2014, blundell_dft_2023}

The isotropic muon hyperfine constant depends on the electron spin density at the location of the muon,\cite{amato_introduction_2024} obtained by taking the zero-separation limit of the muon–electron pair density function (PDF). This dependence exposes a key limitation of DFT methods. Since DFT is a mean-field-like theory that produces one-electron wavefunctions, two-particle correlation functions such as the pair density are not directly available.\cite{perdew_density_2003}
One solution to this problem is to treat the muon as a fixed classical particle in the Born-Oppenheimer approximation, in which case the muon-electron pair density is replaced by the electron density near the fixed muon.\cite{thomas_hyperfine_2006, yamada_accurate_2014} The results of DFT simulations with fixed muons are often good, but the mass of a muon is only 207 times that of an electron (about one ninth that of a proton) and their accuracy is far from guaranteed. In some systems, quantum effects such as zero-point motion are known to influence computed hyperfine constants significantly.\cite{patrignani_review_2016}  Approaches such as path-integral molecular dynamics are often used to capture nuclear quantum effects in a DFT framework and have also been used to sample distributions of muon positions in $\upmu$SR simulations,\cite{yamada_accurate_2014} but the muon-electron pair density is still obtained by treating the muon as fixed at each position.

Two-component DFT, often used to study electron-positron systems, is able to treat both electrons and muons quantum mechanically and can compute the electron and muon densities, but remains unable to describe electron-muon correlations.\cite{boronski_electron-positron_1986} The zero-separation limit of the electron-muon pair density is written as the product of the electron and muon densities multiplied by a correlation factor that must be approximated.\cite{goli_two-component_2022} Local-density-like approximations, in which the correlation factor is assumed to depend on the electron and muon densities (and perhaps their gradients) at the coincidence point, have been used to compute $\upmu$SR frequencies in some cases.\cite{deng_two-component_2023}

Over the past few years, the neural wavefunction approach shown itself able to describe molecules without muons at a level of accuracy comparable to high-level quantum chemistry methods such as coupled cluster with singles, doubles, and perturbative triples (CCSD(T)), whilst scaling more favorably to large system sizes.\cite{glehn_self-attention_2023, pfau_ab_2020, hermann_ab_2023, choo_fermionic_2020, hermann_deep-neural-network_2020} Further, neural-network-based wavefunctions have been used to study quantum systems with multiple species of particles, such as positrons in molecules,\cite{cassella_neural_2024} and dense hydrogen with quantum protons.
\cite{linteau_neural_2025} In this paper, we treat electron-muon correlations quantum mechanically by constructing a neural-network-based many-particle wavefunction ansatz that depends on the positions of electrons and muons, treating both on exactly the same footing. This yields more accurate estimates of muon hyperfine constants than approaches that treat the muon as a classical point particle. We compute the isotropic muon hyperfine constant in the muoniated methyl and ethyl radicals, comparing results obtained when the muon is treated either as a classical point particle or as a quantum particle within the many‑body wavefunction. By comparing the predicted muon hyperfine constants with low-temperature experimental data, we show that a fully quantum treatment of the muon-electron system yields values closer to experiment. Furthermore, when accounting for systematic environmental effects, our approach produces results that are more accurate than standard DFT calculations. We hope that neural wavefunction methods will in future become an essential part of the computational toolkit used to interpret $\upmu$SR experiments.

\subsection{Theory}
We consider a chemical system described to a first approximation by the many-body Coulomb Hamiltonian in the Born-Oppenheimer clamped-nuclei approximation:
\begin{equation}
\begin{aligned}
\label{coloumb_hamiltonian}
\hat{\mathcal{H}}_{\textbf{Coulomb}} = -\sum_{i}  \frac{\hbar^2}{2m_i}\nabla_i^2 + \sum_{ij} \frac{q_i Q_j}{4\pi \epsilon_0|\hat{\mathbf{r}}_i - \mathbf{R}_j|} \\ + \sum_{i>j} \frac{q_i q_j}{4\pi \epsilon_0|\hat{\mathbf{r}}_i - \hat{\mathbf{r}}_j|} + \sum_{i>j} \frac{Q_i Q_j}{4\pi \epsilon_0|\mathbf{R}_i - \mathbf{R}_j|}, 
\end{aligned}
\end{equation}
where $q_i$, $m_i$, and $\hat{\mathbf{r}}_i$ denote the charges, masses, and position operators of the muon and electrons, while $Q_i$ and $\textbf{R}_i$ are the charges and positions of the fixed classical nuclei. 

In addition to the Coulomb interaction, the full Hamiltonian includes hyperfine interactions between the spins of electrons and muons.\cite{amato_introduction_2024} For molecular systems in the gas phase, we only need to consider the isotropic component, also known as the Fermi contact interaction, as anisotropic contributions are averaged out due to the random orientations of the molecules.\cite{mckenzie_analysis_2024} The isotropic hyperfine interaction for many electrons and a single muon can be written as:
\begin{equation}
\label{fermi_contact_hamiltonian}
\hat{\mathcal{H}}_{\textbf{hyperfine}} = \frac{2\mu_0}{3}\gamma_\upe\gamma_{\upmu}\hbar^2 \hat{\mathbf{I}}_\upmu \cdot \sum_{i}^{N}  \hat{\mathbf{S}}_i \delta(\hat{\mathbf{r}}_i - \hat{\mathbf{r}}_\upmu),
\end{equation}
where $\hat{\mathbf{I}}_\upmu$ is the muon spin operator, $\hat{\mathbf{S}}_i$ is the spin operator for the $i$-th electron, and $\hat{\mathbf{r}}_{\upmu}$ and $\hat{\mathbf{r}}_i$ are the corresponding position operators.\footnote{Generalized from Amato. \cite{amato_introduction_2024}} The constant $\mu_0$ is the vacuum permeability, and $\gamma_\upe$ and $\gamma_\upmu$ are the electron and muon gyromagnetic ratios, respectively.

The ground state of the system with a muon and $N$ electrons is described by the many-body wavefunction
\begin{equation}
    \Psi = \Psi(\mathbf{r}_{\upmu}, \sigma_{\upmu};\mathbf{r}_1, \sigma_1; \ldots; \mathbf{r}_N, \sigma_N),
\end{equation}
which depends on the spatial coordinates $\mathbf{r}$ and spin coordinates $\sigma$ of the muon and all of the electrons. It is useful to define the spin-resolved pair density function,
\begin{equation}
    \rho_{\sigma_\upmu \sigma_\upe}(\mathbf{r}_\upmu, \mathbf{r}_\upe) = \sum_{\{\sigma\}} \int d\mathbf{R}_\upe \sum_{i=1}^{N} \delta(\mathbf{r}_\upe - \mathbf{r}_i) \delta_{\sigma_\upe \sigma_i} |\Psi|^2,
\end{equation}
where $\sigma_\upmu$ and $\sigma_\upe$ are the muon and electron spin indices, $d\mathbf{R}_\upe \equiv d^3r_1 \dots d^3r_N$, and $\{\sigma\} \equiv \{ \sigma_1, \dots, \sigma_N \}$. The spin-resolved system-averaged pair density function is then defined as:
\begin{equation}
    \bar{\rho}_{\sigma_{\upmu} \sigma_\upe}(\mathbf{u}) = \int d^3\mathbf{r}_{\upmu} \, \rho_{\sigma_\upmu \sigma_\upe}(\mathbf{r}_{\upmu}, \mathbf{r}_{\upmu} + \mathbf{u}),
\end{equation}
where $\mathbf{u} = \mathbf{r}_\upe - \mathbf{r}_\upmu$ is the separation vector between the muon and the electron. 

Recall the definition of the isotropic muon hyperfine constant $A_\upmu$ (henceforth just \enquote{hyperfine constant}):
\begin{equation}
    \langle \hat{\mathcal{H}}_{\textbf{hyperfine}} \rangle = A_\upmu \langle \hat{\mathbf{I}}_\upmu \cdot \hat{\mathbf{S}} \rangle,
\end{equation}
where $\hat{\mathbf{S}} = \sum_i \hat{\mathbf{S}}_i$ is the total electron spin operator.\cite{amato_introduction_2024}$^{,}$\footnote{The hyperfine constant represents the energy shift due to the alignment or anti-alignment of the muon and system spins.} Placing all particles in $\sigma_z$ eigenstates, parallel or anti-parallel to the applied magnetic field, and comparing the definitions of the Fermi contact Hamiltonian and the spin-resolved system‑averaged pair density, we find that
\begin{equation}
    A_\upmu = \frac{2\mu_0}{3}\gamma_\upe\gamma_{\upmu}\hbar^2 \Delta\rho(0),
\end{equation}
where we have defined
\begin{equation}
    \Delta\rho(\mathbf{r}) = \bar{\rho}_{\uparrow \uparrow}(\mathbf{r}) - \bar{\rho}_{\uparrow \downarrow}(\mathbf{r}),
\end{equation}
and assumed, without loss of generality, that the muon in spin up. 

\section{Method}
Evaluating the hyperfine constant requires finding the ground state of the full Hamiltonian, $\mathcal{\hat{H}} = \mathcal{\hat{H}}_{\textbf{Coulomb}} + \mathcal{\hat{H}}_{\textbf{hyperfine}}$, a task simplified by noting that the Coulomb interaction is several orders of magnitude stronger than the hyperfine interaction. For example, in muonium, the Coulomb ground state energy is $13.54$ eV and the hyperfine constant is $18.5 \times 10^{-6}$ eV --- the latter is approximately 700,000 times smaller than the former.\cite{amato_introduction_2024} As such, first-order perturbation theory can be employed and we need only find the ground state of the Coulombic system, $\mathcal{\hat{H}} = \mathcal{\hat{H}}_{\textbf{Coulomb}}$. We use Hartree atomic units, with energies measured in Hartrees ($E_\text{h}$) and lengths in Bohr radii~($a_0$).

The variational Monte Carlo (VMC) method uses a parameterized trial ground-state wavefunction, $\Psi(\textbf{R};\theta)$, where $\mathbf{R}$ is shorthand for the list of all particle position coordinates, $\mathbf{R} \coloneqq (\mathbf{r}_{\upmu}, \mathbf{r}_1, \mathbf{r}_2, \ldots, \mathbf{r}_N)$, and $\theta$ represents the adjustable parameters. As is always possible when the Hamiltonian is independent of spin, we have chosen to assign the spins,\cite{foulkes_quantum_2001} setting the muon spin up, the first $N^{\uparrow}$ electrons spin up, and the remaining $N^{\downarrow} = N - N^{\uparrow}$ electrons spin down. The wavefunction may then be viewed as a function of the spatial coordinates only. The parameters $\theta$ are optimized using a gradient descent procedure employing the KFAC algorithm \cite{martens_optimizing_2015} to minimize the energy expectation value:
\begin{equation}\label{expectation_energy_eq}
    \langle \mathcal{\hat{H}} \rangle_\theta = \frac{\langle \Psi | \mathcal{\hat{H}} | \Psi \rangle}{\langle \Psi | \Psi \rangle}
    = \frac{\int \Psi^*(\textbf{R};\theta) \mathcal{\hat{H}} \Psi(\textbf{R};\theta) d\textbf{R}}{\int \Psi^*(\textbf{R};\theta) \Psi(\textbf{R};\theta) d\textbf{R}}.
\end{equation}
Monte Carlo integration is used to evaluate the high-dimensional integrals, sampling particle configurations $\mathbf{R}_i$ from the probability density $p \propto \Psi^2$ using the Metropolis-Hastings algorithm.\cite{foulkes_quantum_2001}
The expectation value of the energy and its gradient are computed using the following Monte Carlo estimators:\cite{cassella_neural_2024}
\begin{equation}\label{expectation_energy_mc_eq}
\begin{aligned}
     \langle \mathcal{\hat{H}} \rangle_\theta &= \mathbb{E}_{\textbf{R} \sim p} \left[E_{L}(\textbf{R}; \theta)\right],\\
    \nabla_\theta \langle \mathcal{\hat{H}} \rangle_\theta &= 2\mathbb{E}_{\textbf{R} \sim p}  \left[ \left(E_{L}(\textbf{R};\theta) - \langle \mathcal{\hat{H}} \rangle_\theta \right) \nabla_\theta \log |\Psi(\textbf{R};\theta)| \right] ,
\end{aligned}
\end{equation}
where the symbol $\mathbb{E}_{\textbf{R} \sim p}$ indicates an expectation value over points $\textbf{R}$ sampled from the probability distribution $p$, and $E_{L}(\textbf{R}; \theta)=\mathcal{\hat{H}}\Psi(\textbf{R};\theta)/\Psi(\textbf{R};\theta)$ is the local energy. The accuracy of the VMC method depends on the accuracy with which the trial wavefunction is able to approximate the true ground state.

The neural VMC approach is distinguished from traditional applications of VMC by its use of a neural network as the trial wave function. The network inputs are the particle positions, the output is the value of the wavefunction for those inputs, and the variational parameters are the network weights and biases. Neural networks are such flexible and accurate function approximators that the accuracy of this approach rivals and sometimes exceeds that of traditional quantum chemical methods such as CCSD(T).
The neural wavefunctions used in this work are based on the Psiformer architecture,\cite{glehn_self-attention_2023} with the following functional form:
\begin{equation}
\label{psiformer-equation}
\Psi(\mathbf{R}) 
= e^{J(\mathbf{R})} \sum_{k} \prod_{\chi} 
\det\left[
  \phi_{i}^{k\chi}\left(
    \textbf{r}_{j}^{\chi}; \{\textbf{r}_{/j}^{\chi}\}; \{\textbf{r}_{j}^{/\chi}\}
  \right)
\right],
\end{equation}
where $\chi \in \{\upmu, \upe\}$ indexes the particle species. The wavefunction is a sum of determinants, indexed by the integer $k$, containing many-particle orbitals, $\phi_{i}^{k\chi}\left(\textbf{r}_{j}^{\chi};\{\textbf{r}_{/j}^{\chi}\}, ; \{\textbf{r}_{j}^{/\chi}\}\right)$, which are computed by a transformer-based neural network. The arguments of $\phi_i^{k\chi}$ are the position $\textbf{r}_{j}^{\chi}$ of particle $j$ of species $\chi$, the position of every other particle of the same species, $\{\textbf{r}_{/j}^{\chi}\}$, and the position of every particle not of that species, $\{\textbf{r}_{j}^{/\chi}\}$. The position vectors $\textbf{r}_{j}^{\chi}$ are ordered such that $j$ iterates through all spin-up particles followed by all spin-down particles for particle species $\chi$.
The wavefunction also contains an exponential Jastrow factor, $J(\textbf{R})$, represented by an additional small neural network, that aids in representing short-range particle-particle correlations. By construction, the wavefunction is anti-symmetric with respect to the exchange of identical particles, ensuring that the Pauli exclusion principle is satisfied.\cite{glehn_self-attention_2023} Further details of the neural network architecture are found in Appendix A and Refs.~\onlinecite{glehn_self-attention_2023, pfau_ab_2020, cassella_neural_2024}.

To compute the hyperfine constant, we estimate the spin-resolved system-averaged pair density function $\bar{\rho}_{\sigma_\upmu \sigma_\upe}(r)$ using Monte Carlo integration. The hyperfine constant is proportional to the value at the origin of the translationally averaged muon-electron spin density, $\Delta\rho(0) = \bar{\rho}_{\uparrow \uparrow}(0) - \bar{\rho}_{\uparrow \downarrow}(0)$. We find $\Delta\rho(0)$ by fitting the logarithm $\ln|\Delta\rho(r)|$ as a quadratic polynomial constrained to obey the generalized Kato cusp condition between two particles.\cite{pack_cusp_1966} Further details are provided in Appendix B.

\section{Results}

We study two muonic systems: the muoniated methyl radical (CH$_\text{2}$Mu), which contains 1 muon and 9 electrons, and the muoniated ethyl radical (C$_\text{2}$H$_\text{4}$Mu), which contains 1 muon and 17 electrons. For each system, we calculated the hyperfine constant at different levels of approximation, which we refer to as the classical muon, quantum muon, and full quantum treatments:
\begin{description}
    \item[Classical Muon] The muon, hydrogen nuclei, and carbon nuclei are treated as fixed classical point particles within the Born-Oppenheimer approximation. This is the level of approximation employed in most DFT calculations of hyperfine constants (which in addition approximate the electron–electron correlations using an effective exchange–correlation potential) and is most directly comparable to standard DFT results.
    \item[Quantum Muon] The muon is treated as a quantum particle and included, together with the electrons, in the ground-state wavefunction, while the hydrogen nuclei and carbon nuclei remain clamped. This allows for a more accurate evaluation of the hyperfine constant by explicitly sampling the quantum pair‑density between the muon and the electrons.
    \item[Full Quantum (muoniated methyl radical only)] The muon and the hydrogen nuclei are treated as quantum particles in the ground‑state wavefunction, and only the carbon nucleus is fixed. This enables a complete study of the hyperfine constant that includes the perturbation of the hydrogen nuclei due to the implanted muon.
\end{description}

To position the fixed nuclei, we use DFT-computed geometries from the NIST chemistry WebBook.\cite{linstrom_methyl_nodate, linstrom_ethyl_nodate} When comparing predicted and experimental values, only the magnitude of the hyperfine constant is relevant.\cite{amato_introduction_2024} The relative orientations of the spins of the muon and the unpaired electron in the radical do not affect the ground-state wavefunction of the Coulomb Hamiltonian, which excludes the hyperfine interaction and for which the latter is treated as a perturbation. This allows us to choose both the muon and the unpaired electron to be spin up when solving the Coulomb problem to obtain the unperturbed ground-state wavefunction. 

As a validation of our methodology, we compute the properties of the muonium atom. The quantum muon approach yields an isotropic hyperfine constant of $4469(7)$~MHz and a ground-state energy of $-0.49757(3)$~Hartrees, accurately reproducing the established theoretical and experimental values of $4463$~MHz and $-0.49759$~Hartrees.\cite{eides_hyperfine_2019, kanda_new_2021} Treating the muon as a fixed classical particle results in an energy of $-0.5$~Hartrees and a hyperfine constant of $4529$~MHz,\footnote{Obtained using the electron density at a hydrogen nucleus $\frac{1}{\pi a_0^3}$} as this approximation fails to account for the significant motion of the quantum muon.

\subsection{Muoniated Methyl Radical}

The results of our calculations of the hyperfine constant of the muoniated methyl radical are compared to DFT and experiment in Table \ref{tab:methyl_hyperfine_constants}. The magnitude of the classical-muon hyperfine constant deviates from the experimental magnitude by 27(3) MHz (13\%); this deviation is reduced to 22(3) MHz (11\%) for the quantum-muon calculation, and to 17(3) MHz (8\%) for the full quantum calculation. (Note that the experiments do not determine the \emph{sign} of the hyperfine constant.) Furthermore, the classical-muon value is 40(3) MHz (21\%) larger in magnitude than the DFT prediction, despite describing the same underlying geometry. The fortuitous agreement between classical muon DFT and experiment is discussed further in Sec.~\ref{subsec:discussion-DFT}. 

The corresponding ground-state energies are reported alongside DFT results in Table \ref{tab:methyl_energies}. The agreement between the DFT and classical-muon Psiformer system energies is good, although DFT energies are not variational and may lie below the exact ground-state energy. As more particles are treated quantum mechanically, the system becomes less bound, consistent with the additional zero‑point energy introduced by quantum motion.\cite{moller_quantum_2013} The full quantum Psiformer simulations yield hyperfine constants closest to experiment, reflecting the more realistic description of the relaxation of the hydrogen nuclei in the presence of a quantum mechanical muon. Although the DFT hyperfine constant appears surprisingly accurate, Sec.~\ref{subsec:discussion-environment} shows that it is too small when environmental effects are taken into account.

\begin{table}
\caption{\label{tab:methyl_hyperfine_constants}Isotropic muon hyperfine constants for the muoniated methyl radical computed with the Psiformer and DFT, compared with experimental data at 73 K. Brackets denote uncertainties in the last digit, where reported.}
\begin{ruledtabular}
\begin{tabular}{lc}
Muoniated Methyl Radical & Hyperfine Constant $A_\upmu$ (MHz) \\
\hline
Psiformer Classical Muon & -228(3) \\
Psiformer Quantum Muon & -223(3) \\
Psiformer Full Quantum & -218(3) \\
DFT B3LYP/cc-pVDZ~\cite{pratt_anisotropic_2025} & -188 \\
Experiment (73 K)~\cite{mckenzie_hyperfine_2007} & $\pm$201.30(10)\\
\end{tabular}
\end{ruledtabular}
\end{table}

\begin{table}
\caption{\label{tab:methyl_energies}Ground-state energies for the muoniated methyl radical computed with Psiformer and DFT. Brackets indicate the standard error on the mean of the last digit, where reported.}
\begin{ruledtabular}
\begin{tabular}{lc}
Muoniated Methyl Radical & Ground State Energy ($E_\text{h}$) \\
\hline
Classical Muon & -39.8345(1) \\
Quantum Muon & -39.8012(1) \\
Full Quantum & -39.7814(2) \\
DFT NIST B3LYP/6-31G*~\cite{linstrom_methyl_nodate} & -39.8383 \\
\end{tabular}
\end{ruledtabular}
\end{table}

\begin{figure*}
\includegraphics[width=1\textwidth]{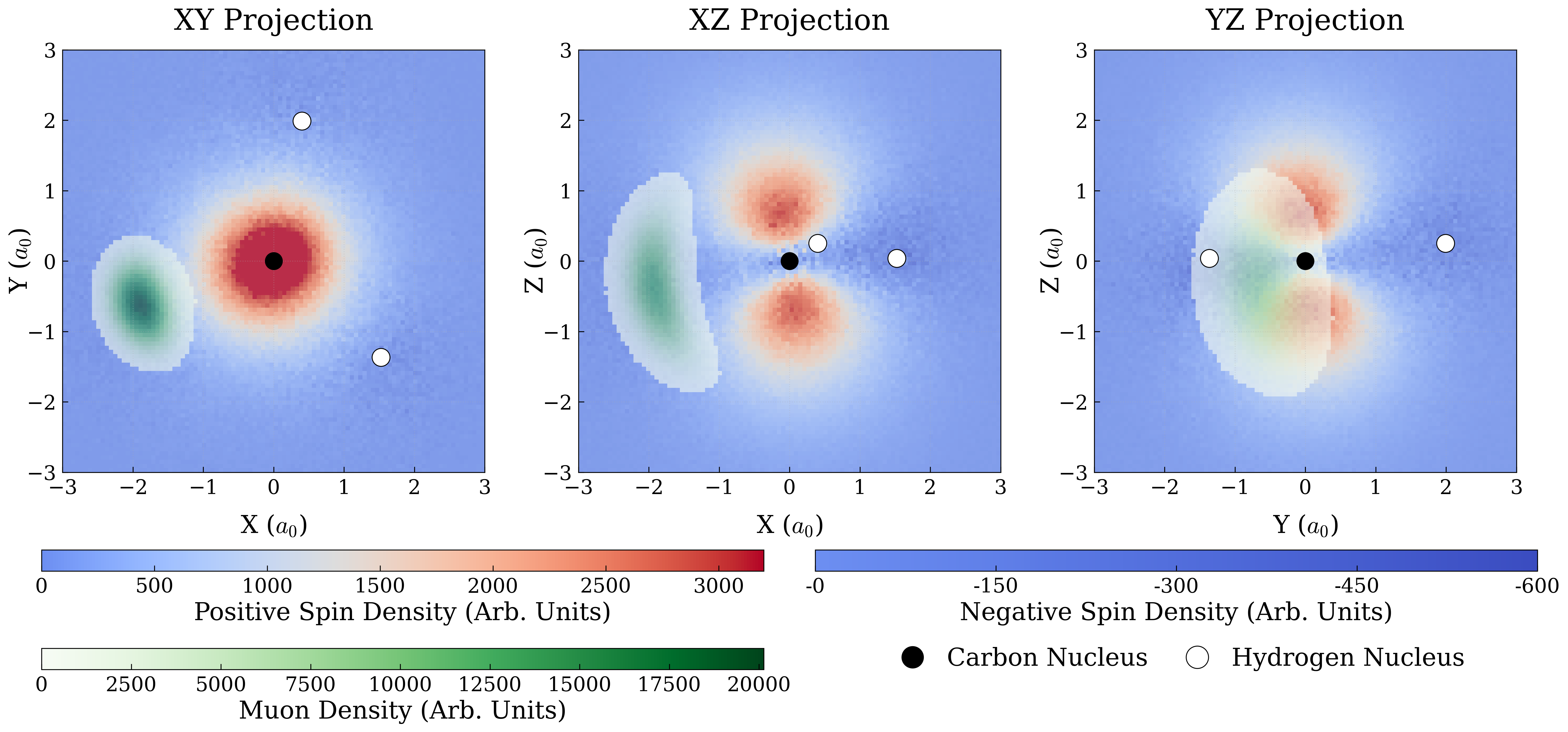}%
\caption{\label{fig:methyl_density_projections}Electron spin density and muon density of the muoniated methyl radical (CH$_\text{2}$Mu) with a quantum muon, projected into the XY, XZ, and YZ planes. The molecular structure is largely coplanar with the XY plane. The magnitude and sign of the electron spin density are indicated by blue-red colors, while the magnitude of the muon density is indicated in green. The electronic spin density is concentrated near the carbon atom, while the muon density is delocalized and centered some distance away.}
\end{figure*}

\begin{figure}
\includegraphics[width=0.48\textwidth]{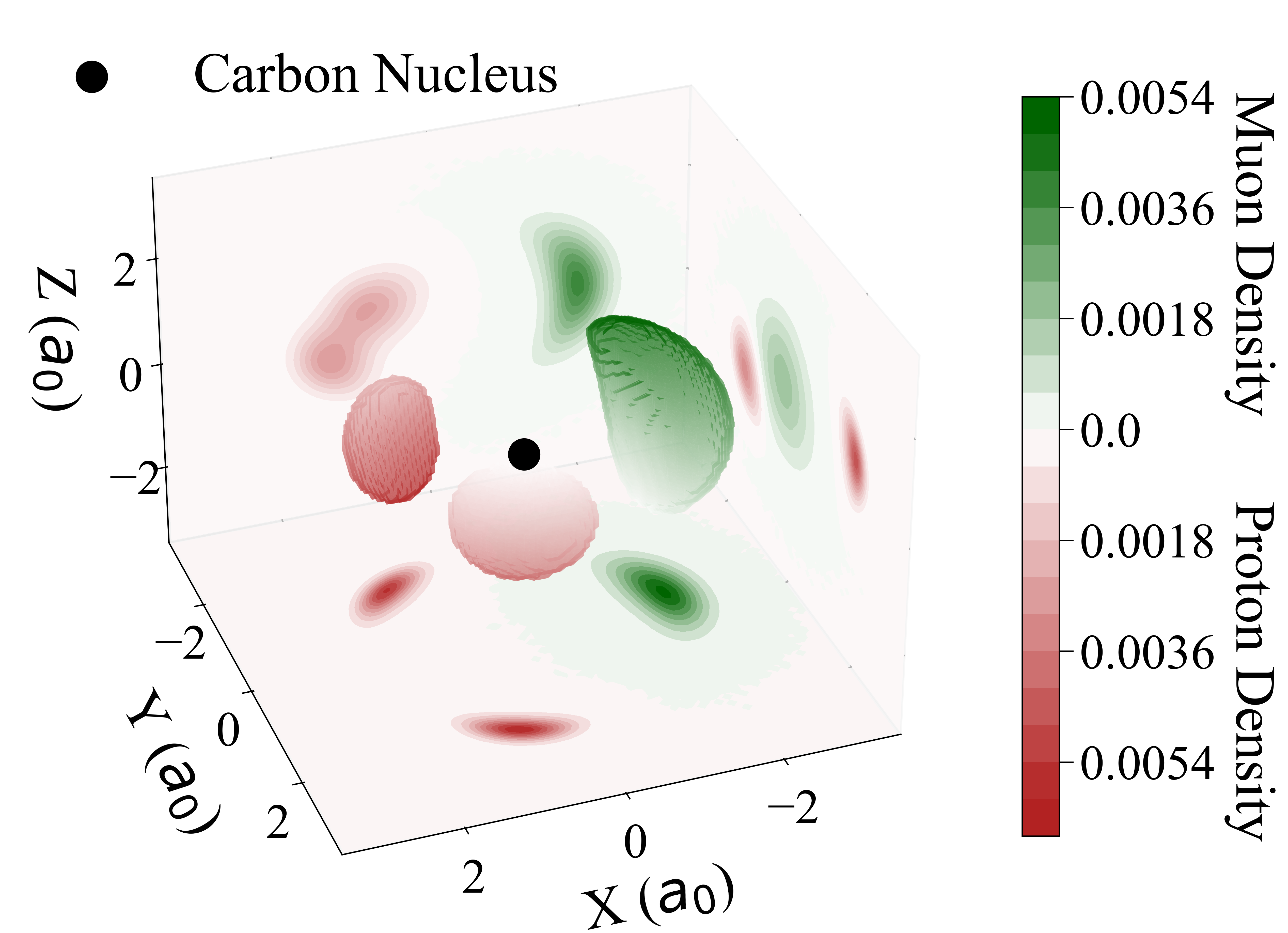}%
\caption{\label{fig:methyl_proton_muon_density}
10\% isosurfaces of the muon and proton densities in the ground state of the muoniated methyl radical (CH$_\text{2}$Mu) obtained from a full quantum Psiformer simulation. The proton density is indicated in red and the muon density in green. The densities were accumulated in $5.12 \times 10^{−4} \;a_0^{\,3}$ cubic bins and smoothed using a Gaussian convolution with $\sigma = 0.06 a_0$. Contour projections of the binned function are shown on the axes. The quantum muon is more delocalized than the protons due to its lower mass. }
\end{figure}

\begin{figure}
\includegraphics[width=0.48\textwidth]{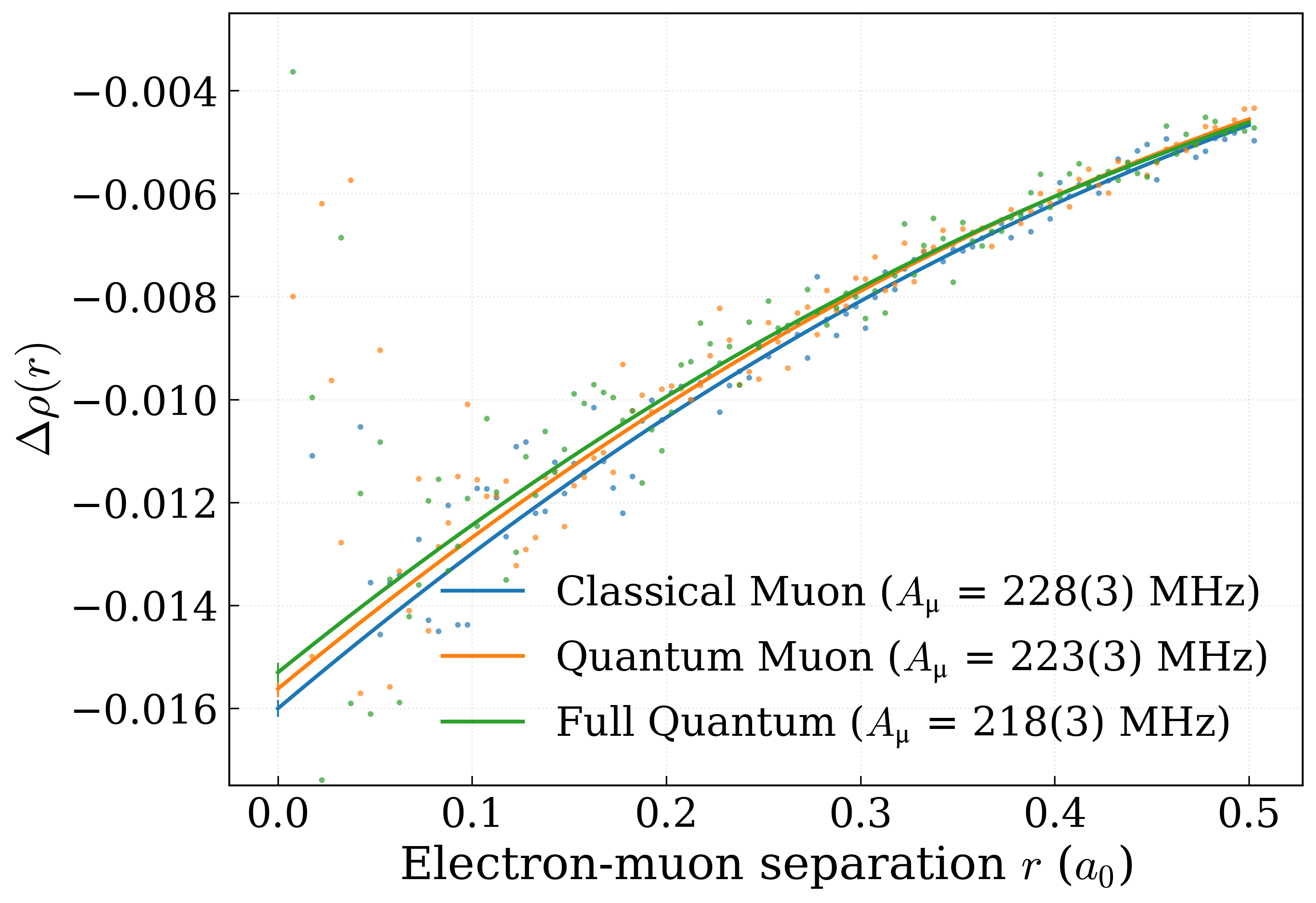}%
\caption{\label{fig:methyl_muon_spin_correlations}Translationally averaged muon-electron spin density for the muoniated methyl radical (CH$_\text{2}$Mu) in the classical muon, quantum muon, and full quantum systems. The muon-electron spin density is extrapolated to obtain $\Delta\rho(r=0)$ using a quadratic fit in the log domain. The extrapolated value and its statistical error determine the isotropic muon hyperfine constant $A_\upmu$ and its uncertainty.}
\end{figure}

Figure \ref{fig:methyl_density_projections} shows planar projections of the muon and electron spin densities when the muon is treated quantum mechanically. The unpaired electron occupies a $p$-like orbital centered on the carbon, but it is the very small negative electron spin density close to the plane of the molecule near the delocalized muon --- hard to discern from the figure alone --- that determines the sign of the calculated muon hyperfine constant. 

The need to treat the muon quantum mechanically is illustrated more clearly in Fig.~\ref{fig:methyl_proton_muon_density}: the muon is noticeably more delocalized than the protons when both are treated quantum mechanically, indicating that the Born-Oppenheimer approximation used in standard DFT may be insufficient to capture the muon’s quantum nature.

Figure \ref{fig:methyl_muon_spin_correlations} shows the sampled translationally averaged muon-electron spin density $\Delta\rho(r)$ as a function of the separation $r$ between the electron and muon. For each system, the sampled electron-muon spacings are placed into radial bins of width 0.005$a_0$ and the number of samples of each spin in each bin divided by the bin volume to obtain an estimate of the sample density and thus the muon-electron spin density at that separation. Every point shows the result of this calculation for one bin in one system. The muon-electron spin density is extrapolated to $r=0$ using a quadratic fit in the log domain, the parameters of which are chosen to ensure that the Kato cusp condition is satisfied exactly. The extrapolated value and its statistical error determine the isotropic muon hyperfine constant $A_\upmu$ and its uncertainty. See Appendix \ref{sec:AppendixB} for details of the data analysis and fitting procedure. All three systems produce similar values for $\Delta\rho(0)$, although the differences between them are greater than the fitting errors.

\begin{figure*}
\includegraphics[width=1\textwidth]{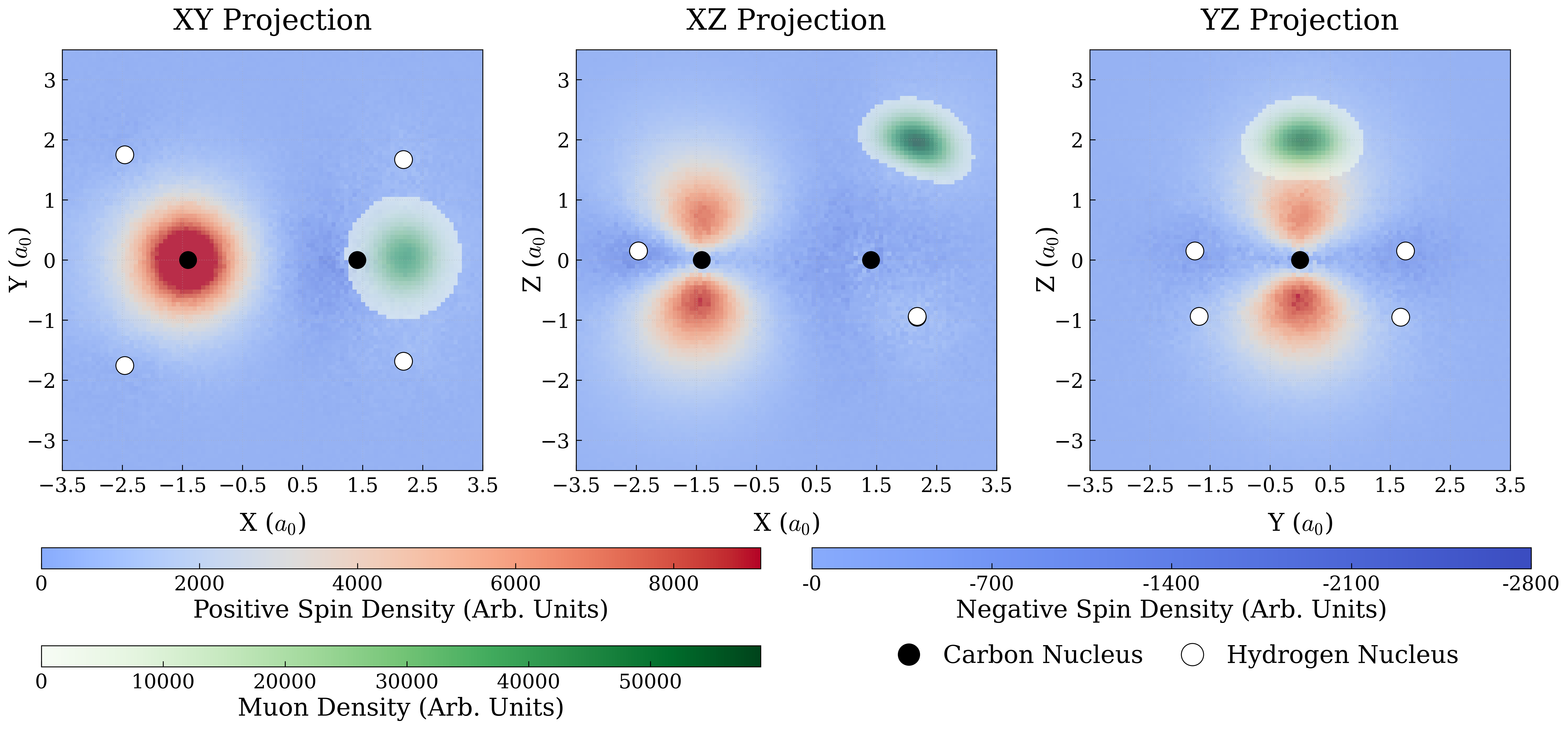}%
\caption{\label{fig:ethyl_density_projections}Electron spin density and muon density of the muoniated ethyl radical (C$_\text{2}$H$_\text{4}$Mu) with a quantum muon, projected into the XY, XZ, and YZ planes. The magnitude and sign of the spin density are indicated by blue-red colors while the magnitude of the muon density is indicated in green. The electronic spin density is concentrated near the carbon atom at $X\approx-1.5 a_0$, while the muon density is delocalized some distance from the other carbon atom at $X\approx1.5 a_0$.}
\end{figure*}

\subsection{Muoniated Ethyl Radical}

\begin{table}
\caption{\label{tab:ethyl_hyperfine_constants}Isotropic muon hyperfine constants of the muoniated ethyl radical computed with the Psiformer and DFT, compared with experimental results. Brackets denote uncertainties in the last digit, where reported.}
\begin{ruledtabular}
\begin{tabular}{lc}
Muoniated Ethyl Radical & Hyperfine Constant $A_\upmu$ (MHz) \\
\hline
Psiformer Classical Muon & 466(4) \\
Psiformer Quantum Muon & 537(4) \\
DFT O3LYP/6-31G(d,p)$^{a,}$~\cite{yamada_accurate_2014} & 478.9 \\
DFT O3LYP/Chipman$^{a,}$~\cite{yamada_accurate_2014} & 493.8 \\
Experiment (5-6 K)$^{b,}$~\cite{bridges_hyperfine_2007} & $\pm$530(3), $\pm$535(2), $\pm$542(2)\\
\end{tabular}
\end{ruledtabular}
\footnotetext{The two DFT results use different basis sets: 6-31G(d,p) and Chipman. The Chipman basis is more accurate as it is designed for Fermi contact spin density calculations.\cite{chipman_gaussian_1989} }
\footnotetext{The muoniated ethyl radical is created in an environment of supercages of USY, HY, and NaY, silicate minerals with porous crystalline structures.~\cite{bridges_hyperfine_2007, ghandi_theoretical_2005}}
\end{table}

\begin{table}
\caption{\label{tab:ethyl_energies}Ground-state energies for the muoniated ethyl radical computed with Psiformer and DFT. Brackets indicate the standard error on the mean of the last digit, where reported.}
\begin{ruledtabular}
\begin{tabular}{lc}
Muoniated Ethyl Radical & Ground State Energy ($E_\text{h}$) \\
\hline
Psiformer Classical Muon & -79.14862(5) \\
Psiformer Quantum Muon & -79.1128(2) \\
NIST DFT B3LYP/6-31G*~\cite{linstrom_ethyl_nodate} & -79.1579 \\
\end{tabular}
\end{ruledtabular}
\end{table}

The results of our calculations of the hyperfine constant of the muoniated ethyl radical are compared with DFT and experiment in Table \ref{tab:ethyl_hyperfine_constants}. The quantum-muon hyperfine constant is consistent with experiment and 71(6) MHz (15\%) larger in magnitude than the hyperfine constant calculated assuming a fixed classical muon. Furthermore, the classical-muon hyperfine constant is smaller in magnitude than the DFT predictions by 13(4) MHz (3\%) or 28(4) MHz (6\%), depending on the DFT basis set.

As shown in Table \ref{tab:ethyl_energies}, the fixed-muon ground-state energies computed using DFT and the Psiformer are in reasonably good agreement. By contrast, when the muon is treated quantum mechanically, the Psiformer energies are less bound, reflecting the additional muon zero‑point energy, as in the methyl radical case.

Figure \ref{fig:ethyl_density_projections} shows planar projections of the muon and electron spin densities for the quantum-muon system. The hyperfine constant is positive because the electron spin density at the muon site, arising from the tail of the singly occupied $p$ orbital on the carbon atom opposite the muon, is itself positive. This indicates that, in the muoniated ethyl radical, the hyperfine constant is dominated by the spin of the unpaired electron in the radical.

\begin{figure}
\includegraphics[width=0.48\textwidth]{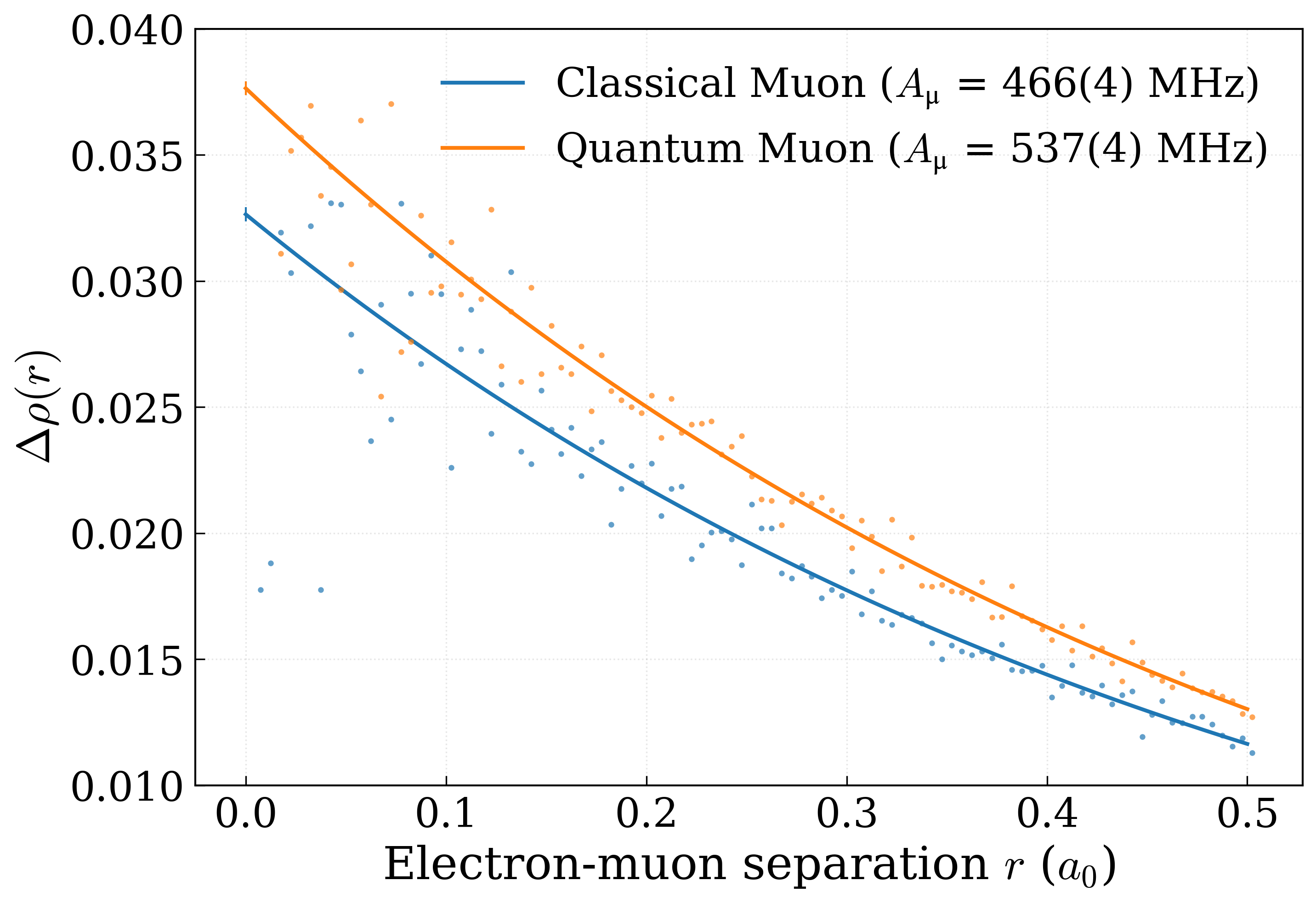}%
\caption{\label{fig:ethyl_muon_spin_correlations}Translationally averaged muon-electron spin densities of the muoniated ethyl radical (C$_\text{2}$H$_\text{4}$Mu) in the classical muon and quantum muon systems. The muon-electron spin density is extrapolated to obtain $\Delta\rho(r=0)$ using a quadratic fit in the log domain. The extrapolated value and its statistical error determine the isotropic muon hyperfine constant $A_{\upmu}$ and its uncertainty.
}
\end{figure}

Figure \ref{fig:ethyl_muon_spin_correlations} shows the sampled translationally averaged muon-electron spin density $\Delta\rho(r)$, accumulated using radial bins of width $0.05 a_0$ and extrapolated to $r=0$ exactly as for the muoniated methyl radical. The difference between the quantum‑muon and classical‑muon calculations can be interpreted as a consequence of muon zero‑point motion, which shifts the muon probability density into regions of higher positive spin density.

\section{Discussions and Conclusions}
\label{sec:discussion}

\subsection{Temperature and Environmental Impacts}
\label{subsec:discussion-environment}

In both the Psiformer and DFT calculations, the hyperfine constant is obtained from ground-state wavefunctions in vacuum at zero temperature. However, temperature and the medium surrounding the muoniated radical can introduce systematic effects that influence experimentally measured hyperfine constants.\cite{ghandi_theoretical_2005, yamada_accurate_2014} 


For the muoniated methyl radical, experimental measurements are available down to 73 K in a liquid ketene environment.\cite{mckenzie_hyperfine_2007} The effect of temperature on the muon hyperfine constant is expected to be small (on the order of 1 MHz) as the results of measurements between 73 K and 223 K vary by no more than 0.34 MHz (0.17\%), and proton hyperfine constants measured between 4 K and 96 K differ by less than 0.3\%.\cite{mckenzie_hyperfine_2007} These observations are consistent with the absence of low-energy vibrational modes that would affect the hyperfine constant below 73 K.

However, proton hyperfine constants for the methyl radical in vacuum at 25 K are larger than in any of the other 11 environments explored by \citet{mckenzie_hyperfine_2007}, irrespective of temperature. We therefore expect the liquid ketene environment to reduce the magnitude of the proton hyperfine constant relative to the vacuum value. By analogy, one would also expect the muon hyperfine constant to reduce. Taking the lowest-temperature measurement for each environment measured by \citet{mckenzie_hyperfine_2007}, the average magnitude of the proton hyperfine constant in methyl is reduced by $1.38 \pm 1.15$ MHz relative to its vacuum value at 25 K, where the uncertainty includes the standard deviation across environments. This corresponds to a fractional reduction of about 2.2\% when moving from vacuum to an average environment.

Assuming an analogous environmental effect for the muon, we can estimate the vacuum value of the muon hyperfine constant by increasing the magnitude of the value in a ketene environment by 2.2\%, from 201.30(10) MHz to about $205.6 \pm 3.6$ MHz. (Since the muon is more delocalized than a proton and likely to be more strongly affected by the environment, the true environmental correction is probably larger.) With this correction, the deviation between experiment and the quantum‑muon Psiformer result is reduced from 22(3) MHz (11\%) to 17(5) MHz (8\%), while the result of the full‑quantum Psiformer calculation differs from experiment by only 12(5) MHz (6\%). The error in the DFT value increases to about 18(4) MHz (9\%).

The experimental measurements for the muoniated ethyl radical were carried out at 5-6 K in silicate mineral environments, which are also expected to affect the muon hyperfine constant.\cite{bridges_hyperfine_2007} The systematic impact of the silicate host can be estimated by comparing room-temperature measurements in these hosts with gas phase data. At 298 K, the gas phase muon hyperfine constant is 329.5(4) MHz,\cite{percival_hyperfine_1989} whereas, at 300 K in NaY, HY, and USY silicate minerals, it is 349(1), 346(1), and 340(1) MHz, respectively,\cite{bridges_hyperfine_2007}. The silicate environment increases the hyperfine constant by 16(4) MHz, on average. 
Experiments also show a strong temperature dependence, making it difficult to correct simultaneously for both environmental and thermal effects; however, at 300 K, the silicate-induced shift is small compared to the overall temperature dependence.
As a simple approximation, we assume that the temperature variation of the environmental offset is small in comparison with the offset itself, and subtract the 16 MHz environmental offset from the low‑temperature silicate data to estimate a corresponding vacuum value. Averaging the 5–6 K measurements in NaY, HY and USY and applying this shift yields an approximate low temperature \emph{vacuum} muon hyperfine constant of 520(5) MHz. The closest DFT prediction underestimates this value by 36(5) MHz (7\%), whereas the quantum muon calculation overestimates it by about 17(6) MHz (3\%).

For both the methyl and ethyl muoniated radicals, the quantum Psiformer calculations of muon hyperfine constants lie closest to the estimated low‑temperature experimental results after environmental effects have been taken into account. However, relying on rough estimates of environmental and thermal corrections is not ideal. In order to more rigorously validate the accuracy of our approach, a direct comparison of the results of Psiformer simulations with low-temperature measurements of solids would be helpful.


\subsection{Comparison to Density Functional Theory}
\label{subsec:discussion-DFT}


Significant deviations exist between Psiformer classical muon and DFT calculations of the muon hyperfine constants of muoniated methyl and ethyl radicals, despite the fact that both approaches treat the muon as a fixed classical particle. In the absence of methodological errors, the two methods would be expected to yield the same hyperfine constant for a given geometry. In DFT, however, the choice of approximate exchange-correlation functional and basis set can also affect the calculated hyperfine constants, as seen for the muoniated ethyl radical in Table \ref{tab:ethyl_hyperfine_constants}.
By contrast, Psiformer employs an \emph{ab initio} variational method that has been shown to represent interacting many-body wavefunctions accurately and to yield highly accurate ground-state energies across a range of systems.\cite{glehn_self-attention_2023, pfau_ab_2020} It is therefore likely that the discrepancies between the fixed-muon hyperfine constants arise primarily from the known limitations of DFT, rather than from a failing of the neural-wavefunction approach. Nonetheless, within these limitations and approximations, the results also illustrate the surprisingly robust performance of DFT.

\subsection{Limitations}

The main limitation of the neural wavefunction method is its cost, which scales as $O(N^{3\text{--}4})$ for $N$ of order several dozen particles. This is not bad in comparison with most other many-body methods:\cite{glehn_self-attention_2023} the cost of a full configuration interaction (FCI) calculation increases exponentially with $N$;\cite{booth_approaching_2010} while standard implementations of the highly accurate CCSD(T) method scale as $O(N^7)$.\cite{glehn_self-attention_2023} However, the pre-factor of the scaling law that determines the cost of a neural wavefunction simulation is large enough to make it impractical to simulate systems with more than around 100 electrons at present.

The cost of conventional implementations of DFT scales as $O(N^3)$. This is not very different from the scaling of neural wavefunction methods, but the DFT prefactor is vastly smaller. For the systems considered here, DFT simulations can be run in seconds on suitable high‑performance computing hardware.\cite{schuch_computational_2009} The most demanding Psiformer calculations performed here --- for the full quantum muoniated methyl radical and the muoniated ethyl radical with a quantum muon ---  required around 7 days on four NVIDIA A100 Tensor Core GPUs. The relative speed of DFT allows nuclear geometries to be relaxed to find optimized structures, which is usually impractical with slower many‑body approaches. A hybrid strategy, in which geometries are optimized with DFT and hyperfine constants are evaluated with a neural‑wavefunction method, may prove effective in future studies.

In the full quantum calculation for the muoniated methyl radical, the proton positions change in response to the presence of the quantum muon. Treating the protons quantum mechanically was not possible in the case of the muoniated ethyl radical due to convergence issues while training the wavefunction. Resolving these convergence issues would provide a scalable method to relax the positions of nearby hydrogen nuclei, and we expect this would yield even more accurate hyperfine constants for a broader range of muoniated free radicals.

Another limitation of the approach is the statistical uncertainty in the calculation of the hyperfine constant. This arises from the limited number of samples accumulated at small electron-muon separations, which leads to the relatively large errors in the on-top translationally averaged muon-electron spin density and the sample deviations observed in Figs \ref{fig:methyl_muon_spin_correlations} and \ref{fig:ethyl_muon_spin_correlations}. This error was mitigated by using a physically motivated fit function constrained by analytical cusp conditions, and performing the fit in the logarithmic domain with weights inversely proportional to the variance to reduce the impact of samples with high variance. Remaining errors could be further reduced by adopting more sophisticated sampling schemes that better resolve the short-range region.

\begin{acknowledgments}

We acknowledge the EUROfusion consortium for providing computing time on the Leonardo Supercomputer at CINECA in Bologna.
The EUROfusion Consortium is funded by the European Union via the Euratom Research and Training Programme
(Grant Agreement No 101052200 — EUROfusion).
Views and opinions expressed are however those of the authors only and do not necessarily reflect those of the European Union or the European Commission.
Neither the European Union nor the European Commission can be held responsible for them.
We also gratefully acknowledge the Gauss Centre for Supercomputing e.V.\ for funding this project
by providing computing time through the John von Neumann Institute for Computing (NIC) on the GCS Supercomputer JUWELS at Jülich Supercomputing Centre.
APF's PhD work is supported by the UK Engineering and Physical Sciences Research Council under grant EP/W524323/1.
Finally, we thank Jarvist Frost, Adrian Hiller, Francis Pratt and Leandro Liborio for helpful discussions.
\end{acknowledgments}

\appendix

\section{Psiformer Hyperparameters}
\label{sec:AppendixA}

All Psiformer calculations were performed using one node and four NVIDIA A100 Tensor Core GPUs of the Booster Module of the Leonardo high-performance computer. The hyperparameters are listed in Table \ref{tab:psiformer_hyperparameters}.

\begin{table}[ht]
\centering
\begin{tabular}{|l|c|}
\hline
\textbf{Parameter} & \textbf{Value} \\
\hline
Psiformer determinants count & $16$ \\
Psiformer layers & $2$ \\
Psiformer heads & $4$ \\
Psiformer heads width & $64$ \\
MLP hidden dims & $256$ \\
Jastrow MLP layers & $3$ \\
Jastrow MLP width & $64$ \\
Learning rate & $  (0.01 + \frac{0.01t}{10000})^{-1}$ \\
Local energy clipping & $5.0$ \\
KFAC Momentum & 0 \\
KFAC Covariance moving average decay & $0.95$ \\
KFAC Norm constraint & $10^{-3}$ \\
KFAC Damping & $10^{-3}$ \\
Steps between parameter updates & 10 \\
Walker batch size & $4096$ \\
Initial walker width & $1.0$ \\
Initial move width & $0.1$ \\
\hline
\end{tabular}
\caption{Psiformer parameters for all systems studied. Learning rate is scheduled based on iteration number $t$}
\label{tab:psiformer_hyperparameters}
\end{table}

\section{Estimation, fitting and statistics of the spin-resolved system-averaged pair density function}
\label{sec:AppendixB}

To compute the hyperfine constant, we estimate the spin-resolved system-averaged pair density function $\bar{\rho}_{\sigma_\upmu \sigma_\upe}(r)$ by accumulating Monte Carlo statistics of particle positions for aligned and anti-aligned electron-muon spin arrangements. Using a histogram approach, we divide the space around the muon into concentric spherical shells, indexed by $k$, with inner radius $r_k$ and outer radius $r_{k+1}$. For $N_\text{samples}$ Monte Carlo configurations, $n_{\sigma_\upe}(k)$ counts the number of electrons with spin $\sigma_\upe$ found within the $k$-th shell centered on the muon. The pair density for a given radial bin is approximated by normalizing this count by the shell volume and the total number of samples:
\begin{equation}
    \bar{\rho}_{\sigma_\upmu \sigma_\upe}(r_k) \approx \frac{n_{\sigma_\upe}(k)}{\frac{4}{3}\pi\left(r_{k+1}^{3} - r_{k}^{3}\right) N_{\mathrm{samples}}}.
\end{equation}
When the muon is treated as a classical fixed particle, the fixed location of the muon is used as the center of the spherical shells. In contrast, for quantum muon simulations, the muon position within each Monte Carlo sample becomes the center, so that the shell radius is the electron-muon spacing.

The hyperfine constant is proportional to $\Delta\rho(0) = \bar{\rho}_{\uparrow \uparrow}(0) - \bar{\rho}_{\uparrow \downarrow}(0)$. However, the vanishing volume of the spherical shells as $r \rightarrow 0$ results in low sample counts and characteristic Poisson noise in estimates of $\bar{\rho}_{\sigma_\upmu \sigma_\upe}(r)$ when $r$ is small. It is computationally infeasible to generate more samples for this region, so this is addressed by fitting the translationally averaged muon-electron spin density $\Delta\rho(r)$ in the short-range region ($r < 0.5 a_0$) and extrapolating to zero. Because the electron–muon pair density exhibits a cusp‑like singularity at coalescence,\footnote{This cusp arises from the Coulomb singularity in the electron–muon interaction, analogous to Kato's cusp condition for electron–nucleus interactions.} we expect $\Delta\rho(r)$ to exhibit exponential behavior at small $r$. Accordingly, we employ a quadratic fit to the logarithm of the translationally averaged muon-electron spin density:
\begin{equation}
    \ln |\Delta\rho(r)| = a + b r + c r^2,
    \label{eq:log_fit_ansatz}
\end{equation}
where $a$, $b$, and $c$ are fitting parameters. The intercept $a$ yields $\ln \Delta\rho(0)$, from which $\Delta\rho(0)$ is recovered.

We constrain the fit by applying the generalized Kato's cusp condition,\cite{pack_cusp_1966} which describes how the many-body wavefunction behaves as any two distinguishable charged particles approach one another. In the case of a  muon and an electron, this reads:
\begin{equation}
    \label{eq:cuspconditions}
    \left. \frac{d \bar{\rho}}{d r} \right|_{r=0} = 2 \mu_{\upmu \upe} q_\upmu q_\upe \bar{\rho}(0),
\end{equation}
where $\mu_{\upmu \upe}$ is the reduced mass of the muon-electron pair and $q_\upmu, q_\upe$ are their charges, in atomic units. Applying this condition to the translationally averaged muon-electron spin density, $\Delta\rho(r)$, fixes the linear coefficient 
$b$ in the logarithmic fitting ansatz,
\begin{equation}
 \left.\frac{\mathrm{d}}{\mathrm{d}r}\ln \Delta\rho(r)\right|_{r=0} = b = -2\mu_{\upmu \upe},
 \label{eq:log_cusp_condition}
\end{equation}
thereby reducing the number of free fitting parameters. A weighted least-squared fit is employed for the remaining parameters, using weights inversely proportional to the variance of each sample. The reported uncertainty in the hyperfine constant is obtained by propagating the standard error of the fitted intercept parameter, $a$, derived from the covariance matrix of the fit. The jackknife and bootstrap procedures\cite{bradley_jackknife_1982} were also used to estimate the uncertainty in the intercept parameter. All three methods gave consistent results, yielding uncertainties of approximately 3 MHz and 5 MHz in the muoniated methyl and ethyl hyperfine constants, respectively.

Two validity checks were employed for all results. First, we verified that the integral of $\Delta\rho(\mathbf{r})$ was unity to within floating-point error, consistent with the presence of a single unpaired electron in each system. Second, although the fit for $\Delta\rho(r)$ explicitly imposes the Kato cusp condition, the neural network does not. The high quality of the resulting fits suggests that the neural network has learned an accurate representation of the cusp.


%
%



\bibliography{references}

\end{document}